\documentclass{aa}
\usepackage{graphicx}
\usepackage{txfonts}
\begin{document}
   \title{The fundamental parameters of the roAp star $\gamma$ Equulei}

   \author{K. Perraut\inst{1}, I. Brand\~ao\inst{2}, D. Mourard\inst{3},
M. Cunha\inst{2}, Ph. B\'{e}rio\inst{3}, D. Bonneau\inst{3}, O. Chesneau\inst{3}, J.M.~Clausse\inst{3},
O. Delaa\inst{3}, A. Marcotto\inst{3}, A. Roussel\inst{3}, A. Spang\inst{3}, Ph. Stee\inst{3}, I. Tallon-Bosc\inst{4},
H. McAlister\inst{5,6}, T.~ten~Brummelaar\inst{6}, J. Sturmann\inst{6}, L. Sturmann\inst{6}, N. Turner\inst{6},
C. Farrington\inst{6} and P.J. Goldfinger\inst{6}}

   \institute{Laboratoire d'Astrophysique de Grenoble (LAOG),
         Universit\'e Joseph-Fourier, UMR 5571 CNRS, BP 53, 38041
         Grenoble Cedex 09, France
         \and
             Centro de astrof\'\i sica e Faculdade de Ci\^encias, Universidade do Porto, Portugal
             \and
             Laboratoire Fizeau, OCA/UNS/CNRS
         UMR6525, Parc Valrose, 06108 Nice cedex 2, France
             \and
             Universit\'{e} de Lyon, Lyon, F-69003, France; Universit\'{e}
         Lyon~1, Observatoire de Lyon, 9 avenue Charles Andr\'{e},
         Saint Genis Laval, F-69230; CNRS, UMR 5574, Centre de
         Recherche Astrophysique de Lyon; Ecole Normale
         Sup\'{e}rieure, Lyon, F-69007, France
         \and Georgia State University, P.O. Box 3969, Atlanta GA
         30302-3969, USA
         \and CHARA Array, Mount Wilson Observatory,
         91023 Mount Wilson CA, USA
             }
\offprints{Karine.Perraut@obs.ujf-grenoble.fr}
   \date{Received ...; accepted ...}

     \abstract
   {Physical processes working in the stellar interiors as well as the evolution of stars
   depend on some fundamental stellar properties, such as mass, radius, luminosity, and chemical
   abundances. The effective temperature, the surface gravity and the mean density are useful
   quantities defined from these fundamental properties. Additional physical quantities, like
   mass loss rate, pulsation period, rotation period, and magnetic field properties are interesting
   for the study of peculiar evolutionary stages. A classical way to test stellar interior models
   is to compare the predicted and observed location of a star on theoretical evolutionary tracks
   in a H-R diagram. This requires the best possible determinations of stellar mass, radius, luminosity and abundances.}
   {To directly and accurately determine its angular diameter and thus derive its fundamental parameters,
   we observed the well-known rapidly oscillating Ap star, $\gamma$ Equ, using the visible
   spectro-interferometer VEGA installed on the optical CHARA array.}
   {We recorded data on the W1W2 baseline of the CHARA array in the blue and in the red domains. We
   computed the calibrated squared visibility and derived the uniform-disk angular diameter and the
   limb-darkened one. We used the whole energy flux distribution, the parallax and the angular diameter
   to determine the luminosity and the effective temperature of the star.}
   {We obtained a limb-darkened angular diameter of 0.564~$\pm$~0.017~mas and deduced a radius of
   $R$~=~2.20~$\pm$~0.12~${\rm R_{\odot}}$. Without considering the multiple nature of the system, we derived
   a bolometric flux of $(3.12\pm 0.21)\times 10^{-7}$ erg~cm$^{-2}$~s$^{-1}$ and an effective temperature of
   7364~$\pm$~235~K, which is below the effective temperature that has been previously determined. Under the same conditions
   we found a luminosity of $L$~=~12.8~$\pm$~1.4~${\rm L_{\odot}}$. When the contribution of the closest
   companion to the bolometric flux is considered, we found that the effective temperature and luminosity of the
   primary star can be, respectively, up to $\sim$~100~K and up to $\sim$~0.8~L$_\odot$ smaller than the values
   mentioned above.}
   {For the first time, thanks to the unique capabilities of VEGA, we managed to constrain the angular
   diameter of a star as small as 0.564 mas with an accuracy of about 3\%, and to derive its fundamental
   parameters. In particular the new values of the radius and effective temperature should bring further
   constraints on the asteroseismic modelling of the star.}

    \authorrunning {K. Perraut et al.}
   \titlerunning{The fundamental parameters of the roAp $\gamma$ Equ}
   \maketitle

\section{Introduction}

Rapidly oscillating Ap (roAp) stars are chemically peculiar main-sequence stars that are characterized
by strong and large-scale organized magnetic fields (typically of several kG, and up to 24 kG), abundance
inhomogeneities leading to spotted surfaces, small rotational speeds, and pulsations with periods of a few
minutes (see, \cite{kochukhov09,Cunha07}, for recent reviews). roAp stars are bright, pulsate with large
amplitudes and in high radial orders. Thus they are particularly well-suited for asteroseismic campaigns
and they contribute in a unique way to our understanding of the structure and evolution of stars. However,
to put constraints on the interior chemical composition, the mixing length parameter, and the amount of
convective overshooting, asteroseismic data should be combined with high precision stellar radii
(\cite{Cunha03,Cunhaetal07}). This radius is generally estimated from the star's luminosity and effective
temperature. But systematic errors are likely to be present in this determination due to the abnormal surface
layers of the Ap stars. This well known fact has been corroborated by seismic data on roAp stars (\cite{matthews99}),
and compromises all asteroseismic results for this class of pulsators. Using long-baseline interferometry to
provide accurate angular diameters appears to be a promising approach to overcome the difficulties in deriving
accurate global parameters of roAp stars, but is also very challenging due to their small angular size. In fact,
except for $\alpha$~Cir, whose diameter is about 1 millisecond of arc (mas) (\cite{Bruntt}), all roAp stars
have angular diameters smaller than 1~mas. Such a small scale can be resolved only with optical or near-infrared
interferometry. This was confirmed again recently by the interferometric study of the second largest (in angular
size) roAp star known, namely $\beta$~CrB (\cite{Bruntt10}).

$\gamma$ Equ (HD201601~;~A9p~;~$m_{V}$~=~4.7~;~$\pi_P$~=~27.55~$\pm$~0.62~mas (\cite{Hip2007})~;~v~$\sin i \sim$~10~km/s
(Uesugi \& Fukuda 1970)) is one of the brightest objects of the class of roAp stars with a period of about 12.3 min
(Martinez et al. 1996) in brightness as well as in radial velocity. Despite photometry and spectroscopy of its
oscillations obtained over the past 25 years, the pulsation frequency spectrum of $\gamma$ Equ has remained
poorly understood. High-precision photometry with the MOST satellite has led to unique mode identifications
based on a best model (\cite{Gruberbauer08}) using a mass of 1.74 $\pm$ 0.03~M$_{\odot}$, an effective temperature
of log $T_{\rm eff}$~=~3.882~$\pm$~0.011 and a luminosity of log $L/{\rm L_\odot}$~=~1.10~$\pm$~0.03 (\cite{Koch06}).
As regards to abundance inhomogeneities, Ryabchikova et al. (2002) considered the following stellar parameters
($T_{\rm eff}$~=~7700~K, log~$g~=~$4.2, [M/H]~=~+0.5) to compute synthetic spectra and presented the evidence for
abundance stratification in the atmosphere of $\gamma$ Equ: Ca, Cr, Fe, Ba, Si, Na seem to be overabundant in
deeper atmospheric layers, but normal to underabundant in the upper layers, which according to the authors agrees
well with diffusion theory for Ca and Cr, developed for cool magnetic stars with a weak mass loss of about 2.5
$\times 10^{-15}$ M$_{\odot}$/yr. Pr and Nd from the rare earth elements have an opposite profile since their
abundance is more than 6 dex higher in the upper layers than in the deeper atmospheric ones. Such abundance
inhomogeneities clearly lead to a patchy surface, a redistribution of the stellar flux, and a complex atmospheric
structure, resulting in biased photometric and spectroscopic determinations of the effective temperature.

Guided by these considerations, we have observed $\gamma$ Equ with a spectro-interferometer operating at optical
wavelengths, the  VEGA spectrograph (\cite{vega}) installed at the CHARA Array (\cite{chara}). The unique combination
of the visible spectral range of VEGA and the long baselines of CHARA has allowed us to record accurate squared
visibilities at high spatial frequencies (Sect.~2). To derive the fundamental parameters of $\gamma$ Equ, calibrated
spectra have been processed to estimate the bolometric flux and to determine the effective temperature (Sect.~3).
Finally, we can set the star $\gamma$ Equ in the HR diagram and discuss the derived fundamental parameters (Sect.~4).

\section{Interferometric observations and data processing\label{inter}}

\subsection{Data}

Data were collected at the CHARA Array with the VEGA
 spectropolarimeter recording spectrally dispersed fringes at
 visible wavelengths thanks to two photon-counting detectors.
 Two telescopes along the W1W2 baseline were combined.
 Observations were performed between 570 and 750~nm (according to the detector)
 at the medium spectral resolution of VEGA (R = 5000). Observations of $\gamma$ Equ were sandwiched with those
 of a nearby calibration star (HD 195810). The observation log is given in Table~\ref{tab:log}.

\begin{table}[t]
\centering
\caption{Journal of $\gamma$ Equ observations on July 29, and August 3 and 5, 2008.}
\label{tab:log}
\begin{tabular}{ccccc}
\hline
Date & UT (h) & Star & B (m) & PA ($^\circ$) \\
\hline
2008-07-29 & 5.59 & HD 195810 & 78.9 & 106.6 \\
2008-07-29 & 6.08 & $\gamma$ Equ & 76.2 & 106.4 \\
2008-07-29 & 6.41 & HD 195810 & 92.3 & 101.9 \\
\hline
2008-08-03 & 8.64 & HD 195810 & 107.3 & 93.0 \\
2008-08-03 & 8.98 & $\gamma$ Equ & 107.8 & 93.8 \\
2008-08-03 & 9.31 & HD 195810 & 103.7 & 91.0 \\
\hline
2008-08-05 & 7.68 & HD 195810 & 107.3 & 108.8 \\
2008-08-05 & 8.14 & $\gamma$ Equ & 106.7 & 95.8 \\
2008-08-05 & 8.63 & HD 195810 & 106.9 & 92.6 \\
\hline
\end{tabular}
\end{table}

Each set of data was composed of observations following a
calibrator-star-calibrator sequence, with 10 files of 3000 short exposures of 15 ms per
observation. Each data set was processed in 60 files of 500 short exposures
using the $C_{1}$ estimator and the VEGA data
reduction pipeline detailed in Mourard et al. (2009). The spectral separation
between the two detectors is fixed by the optical design and equals about 170 nm
in the medium spectral resolution. The red detector was centered around 750 nm on
July 29 and around 640 nm on August 3 and 5. The blue detector was centered around
590 nm on July, 29 and around 470 nm on August 3 and 5. The bluer the wavelength,
the more stringent the requirements on seeing. As a consequence the blue data on
August 3 and 5 did not have a sufficient signal-to-noise ratio and squared visibilities
could not be processed. All the squared visibilities are calibrated using an
uniform-disk angular diameter of 0.29 $\pm$ 0.02 mas in the V and R bands for the calibrator
HD~195810. This value is determined from the limb-darkened angular diameter provided by
SearchCal\footnotemark{}\footnotetext{$\textrm{http://www.jmmc.fr/searchcal\_page.htm}$}(Table~\ref{tab:V2}).

\begin{table}[t]
\caption{Calibrated squared visibilities of $\gamma$ Equ. Each visibility point corresponds to the average on the 60 blocks of 500 frames.}
\centering
\begin{tabular}{cccc}
\hline
UT (h) & B (m) & $\lambda_0$ (nm) & $V^{2}$\\
\hline
6.08 & 76.1 & 745.0 & 0.84 $\pm$ 0.02 \\
6.08 & 76.2 & 582.5 & 0.72 $\pm$ 0.02 \\
8.98 & 107.6 & 640.0 & 0.62 $\pm$ 0.04 \\
8.14 & 106.7 & 640.0 & 0.61 $\pm$ 0.05 \\
\hline
\label{tab:V2}
\end{tabular}
\end{table}

\subsection{Angular diameter determination}

\begin{figure}[h]
\centering
 \includegraphics[width=9cm, angle=0]{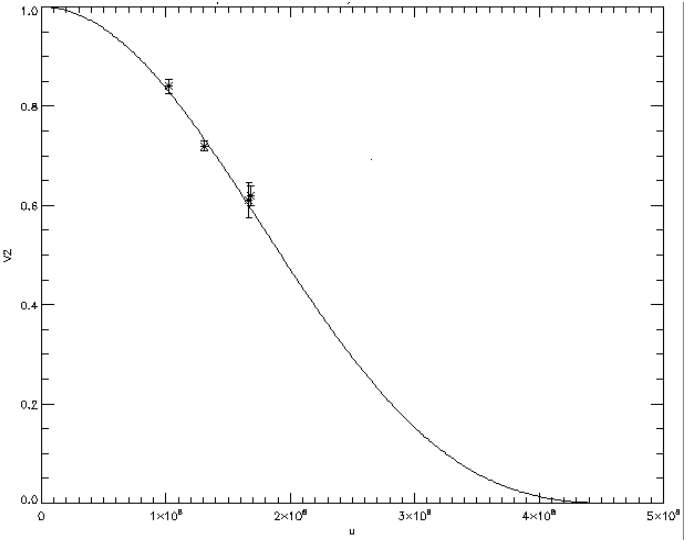}
\caption{Squared visibility versus spatial frequency $u$ for $\gamma$ Equ obtained with the VEGA observations. The solid line
represents the uniform-disk best model.}
  \label{fig:VvsLambda}
\end{figure}

$\gamma$ Equ is the brightest component of a multiple system. The clo\-sest component lies at
1.25"~$\pm$~0.04", it has a magnitude difference with the primary star of $\Delta m$ = 4 and
a position angle of PA~=~264.6$^\circ$~$\pm$~1.3$^\circ$ (\cite{fabricius02}). The entrance
slit of the spectrograph (height=4'' and width=0.2'' for these observations) will affect the
transmission of the companion flux. Taking into account the seeing during the observations
(about 1"), the field rotation during the hour angle range of our observations ([-30$^\circ$ ; 0$^\circ$]),
the position angle of the companion, we determine the throughput efficiency of the VEGA spectrograph
slit for this companion. This efficiency varies from 10\% for the longer baselines (around 107 m)
to 30\% for the smaller ones (around 80 m). We use the Visibility Modeling Tool
(VMT)\footnotemark{}\footnotetext{$\textrm{http://www.nexsciweb.ipâc.caltech.edu/vmt/vmtWeb}$} to
build a composite model including the companion of $\gamma$ Equ. For the longer baselines, the
resulting modulation in the visibility is below 2\%, which is 3 or 4 times below our accuracy
on squared visibilities. We thus neglected the influence of the companion and interpreted our
visibility data points in terms of angular diameter (Fig.~\ref{fig:VvsLambda}). We performed
model fitting with LITpro\footnotemark{}\footnotetext{$\textrm{http://www.jmmc.fr/litpro\_page.htm}$}.
This fitting engine is based on a modified Levenberg-Marquardt algorithm combined with the trust regions
method (\cite{TallonBosc}). The software provides a user-expandable set of geometrical elementary
models of the object, combinable as building blocks. The fit of the visibility curve versus spatial
frequency leads to a uniform-disk angular diameter of 0.540 $\pm$ 0.016 mas for $\gamma$ Equ. We used
the tables of Diaz-Cordoves et al. (1995) to determine the linear limb-darkening coefficient in the R
band for 4.0~$\leq \log g \leq$~4.5 and 7500~K~$\leq T_{\rm eff} \leq$~7750~K. By fixing this limb-darkening
coefficient, LITPRO provides a limb-darkened angular diameter in the R band of $\theta_{LD}$~=~0.564~$\pm$~0.017~mas
with a reduced $\chi^2$ of 0.37.

\section{Bolometric flux and effective temperature}

The effective temperature, $T_{\rm eff}$, of a star can be obtained through the relation,

\begin{equation}\label{eq1}
\sigma T_{\rm eff}^{4} = 4 f_{bol}/\theta_{LD}^{2},
\label{eqteff}
\end{equation}
where $\sigma$ stands for the Stefan-Boltzmann constant ($5.67\times\,10^{-5}$~erg~cm$^{-2}$~s$^{-1}$~K$^{-4}$),
$\theta_{LD}$ for the limb-darkened angular dia\-meter, and $f_{bol}$ is the star's bolometric flux given by,

\begin{equation}
 f_{bol}={\int\limits_0^\infty\,F(\lambda)\mathrm{d}\lambda}.
\label{eqfbol}
\end{equation}

Thus, the effective temperature of $\gamma$ Equ can be computed if we know its angular diameter and its bolometric
flux. The angular diameter of $\gamma$ Equ was derived in Sect.~\ref{inter}. To compute the bolometric flux we
need a single spectrum that covers the whole wavelength range. This spectrum was obtained by combining
photometric and spectroscopic data of $\gamma$ Equ available in the literature, together with ATLAS9 Kurucz
models, in the way explained below.

\subsection{Data}

We collected two rebinned high resolution spectra ($R$~=~18000 at $\lambda$~=~1400~\AA, $R$~=~13000 at
$\lambda$~=~2600~\AA) from the Sky Survey Telescope obtained at the \textit{IUE} ``Newly Extracted Spectra'' (INES)
data archive\footnote{http://sdc.laeff.inta.es/cgi-ines/IUEdbsMY},
covering the wavelength range [1850~\AA~ ; 3350~\AA]. The two spectra were obtained with the Long Wavelength
Prime camera and the large aperture of 10" $\times$ 20" (Table~\ref{tab:iue}).
Based on the quality flag listed in the IUE spectra (\cite{gar97})
we removed all bad pixels from the data, and we also removed
the points with negative flux. The mean of the two spectra was
then computed to obtain one single spectrum of $\gamma$ Equ in the range
1850~\AA~$< \lambda <$~3350~\AA.

\begin{table}
\caption{UV spectra obtained with IUE.}
\label{tab:iue}
\begin{tabular}{cccc}
\hline
Image  &  Date & Starting time & Exposure time\\
Number & & (UT) & (s) \\
\hline
06874 & 08/10/1985 & 18:55:04 & 599.531 \\
09159 & 23/09/1986 & 20:41:13 & 539.730\\
 \hline
\end{tabular}
\end{table}

We collected two spectra for $\gamma$ Equ in the visible, one from Burnashev (1985), which is a spectrum from
Kharitonov et al. (1978) reduced to the uniform spectrophotometric system of the ``Chilean Catalogue'',
and one from Kharitonov et al. (1988). We verified that the latter was in better agreement with the Johnson
(\cite{Morel78}) and the Geneva (\cite{Rufener88}) photometry than the other spectrum. To convert from Johnson
and Geneva magnitudes to fluxes we used the calibrations given by Johnson (1966) and Rufener \& Nicolet (1988),
respectively.

For the infrared, we collected the photometric data available in the literature.
The calibrated observational photometric fluxes that we considered
in this study are given in Table\,\ref{tab:photir}.

\begin{table}
\caption{Calibrated photometric infrared fluxes for $\gamma$ Equ.}
\label{tab:photir}
\begin{minipage}{0.5\textwidth}
\begin{tabular}{ccccc}
\hline
Band & $\lambda_{\rm eff}$ & Flux & Source & Calibration \\
& (\AA) & ($\times10^{-12}\, \rm {erg\, cm^{-2}\,s^{-1}\,\AA^{-1}}$) &  &\\
\hline
I & 9000  & 15.53 & 1 & a
\\
J & 12500 & 5.949 & 2 & b
\\
H & 16500 & 2.420 & 2 & b
\\
K & 22000 & 0.912 & 2 & b
\\
L & 36000 & 0.140 & 2 & b
\\
M & 48000 & 0.0512 & 2 & b
\\
J & 12350 & 6.090 & 3 & c
\\
H & 16620 & 2.584 & 3 & c
\\
K & 21590 & 1.067 & 3 & c
\\ \hline
\end{tabular}
\begin{flushleft}
Source references: (1)\,Morel \& Magnenat (1978); (2)\,Groote \& Kaufmann (1983); (3)\, Cutri et al. (2003).

Calibration references: (a)\,Johnson (1966); (b)\,Wamsteker (1981); (c)\,Cohen et al. (2003).
\end{flushleft}
\end{minipage}
\end{table}

\begin{figure*}[t]
\centering
 \includegraphics[width=18cm, angle=0]{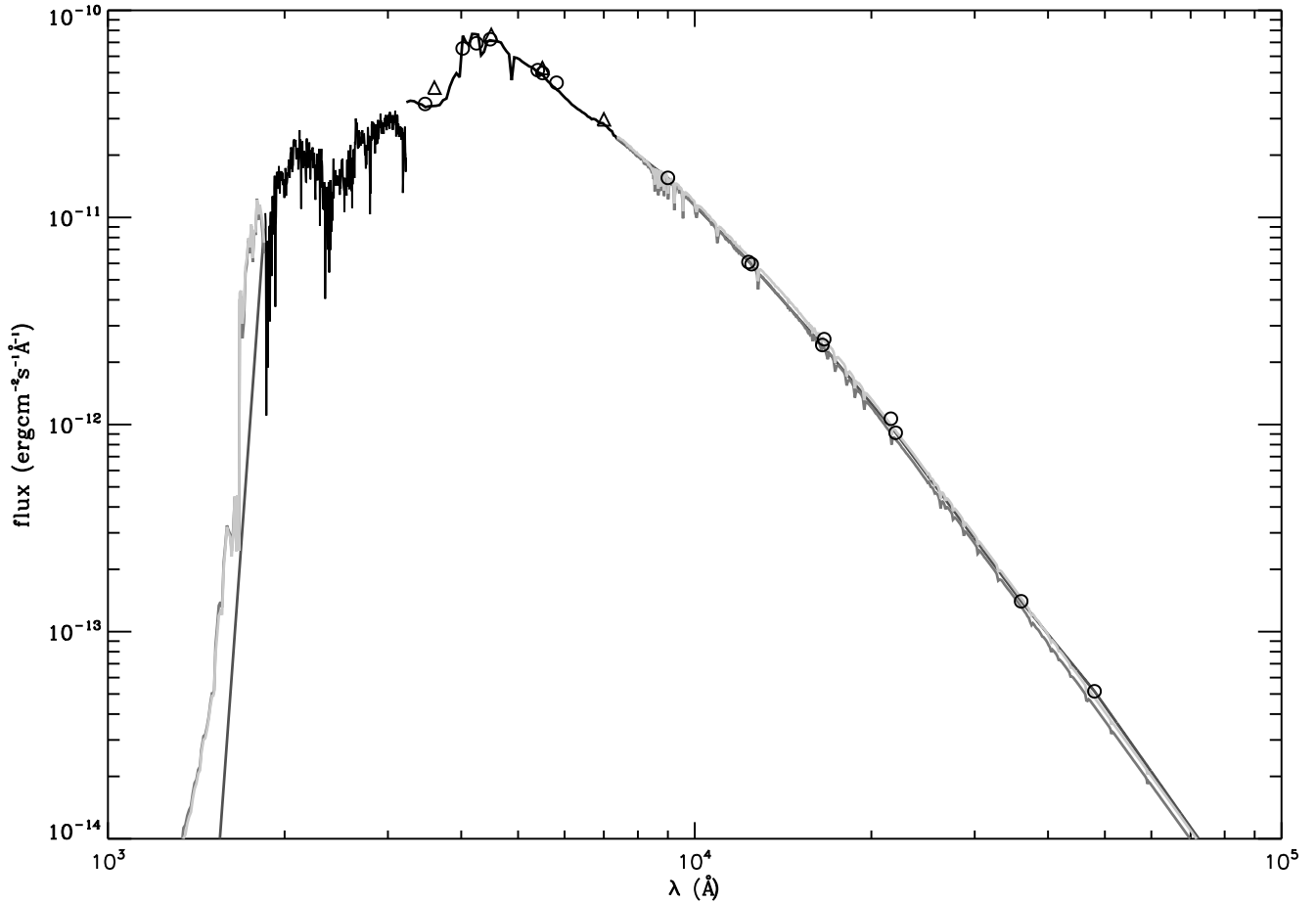}
\caption{The whole spectrum obtained for $\gamma$ Equ. Black line corresponds to the average of the IUE spectra and to the \cite{Kharitonov}'s spectrum. For wavelengths $\rm \lambda < 1854$~\AA~and $\rm \lambda > 7390$~\AA~, the figure shows the curve obtained using the interpolation method (dark grey line), the Kurucz model that best fits the spectroscopy in the visible and the photometry in the infrared when models are calibrated with the star's magnitude $m_{V}$ (grey line) and when models are calibrated with the relation $(R/d)^{2}$ (light grey line). The Geneva and infrared photometry from Table \ref{tab:photir} (circles) and Johnson UBVRI photometry (triangles) are overplotted to the spectrum.}
  \label{fig:plots}
\end{figure*}

\subsection{$f_{bol}$ and $T_{\rm eff}$ determination}

The spectrum of $\gamma$ Equ was obtained by combining the averaged IUE spectrum
between 1854~\AA~ and 3220~\AA, the Kharitonov's (1988) spectrum
from 3225~\AA~to 7375~\AA, and, for wavelengths $\rm \lambda < 1854$~\AA~and
$\rm \lambda > 7390$~\AA~we considered two cases: (1) we used
the synthetic spectrum for the Kurucz model that best fitted
both the star's spectrum in the visible and the star's photometry in
the infrared and, (2) we performed a linear extrapolation between 506\,\AA~and
1854\,\AA,~considering zero flux at 506\,\AA, a second linear interpolation
to the infrared fluxes between 7390\,\AA~and 48000\,\AA,~and a third linear extrapolation
from 48000\,\AA~and 1.6~$\times 10^6$\,\AA~considering zero flux
at 1.6~$\times 10^6$\,\AA.  In case (1), when searching for the best Kurucz model we intentionally
disregarded the data in the UV, because Kurucz models are particularly
unsuitable for modeling that region of the spectra of roAp stars.
To find the Kurucz model that best fitted
the data in the visible and infrared we ran a grid of models,
with different effective temperatures,
surface gravities, and metallicities.
Since Kurucz models needed to be calibrated (they give
the flux of the star, not the value observed on Earth),
we tried two different calibrations, namely:
(i) the star's magnitude in the $V$ band, $m_{V}$,
(ii) the relation $(R/d)^{2}$,
where $R$ is the radius and $d$ the distance to the star.
For the $R/d = \theta/2$ we used the limb-darkened angular diameter $\theta_{LD}$ determined in the previous section.
The final spectra obtained for $\gamma$ Equ with the two different
calibration methods and with the interpolation method
are plotted in Fig.~\ref{fig:plots}. The bolometric flux, $f_{bol}$,
was then computed from the integral of the spectrum of the star,
through Eq.~\ref{eqfbol} and the effective temperature, $T_{\rm eff}$,
was determined using Eq.~\ref{eqteff} (Table~\ref{tab:res}).

\begin{table}[h]
\caption{Bolometric flux $f_{bol}$ and effective temperature $T_{\rm eff}$ obtained for $\gamma$ Equ, using three different methods (see text for details).}
\centering
\begin{tabular}{ccc}
\hline
Calibration method & $f_{bol}$ (erg~cm$^{-2}$~s$^{-1}$) & $T_{\rm eff}$ (K) \\
\hline
$m_{V}$ & $(3.09 \pm 0.20)\times$10$^{-7}$ & 7351 $\pm$ 229\\
$(R/d)^{2}$ & $(3.15 \pm 0.21)\times$ 10$^{-7}$ & 7381 $\pm$ 234\\
\hline
Interpolation & $(3.11 \pm 0.21)\times$ 10$^{-7}$ & 7361 $\pm$ 235\\
\hline
\end{tabular}
\label{tab:res}
\end{table}

The uncertainties in the three values of the bolometric flux given in Table~\ref{tab:res} were estimated by
considering an uncertainty of $10\%$ on the total flux from the combined IUE
spectrum (\cite{Riestra01}), an uncertainty of $4\%$ on the total
flux of the low resolution spectrum from Kharitonov et al. (1988),
an uncertainty of $20\%$ on the total flux derived from the Kurucz model, and
an uncertainty of $20\%$ on the total flux derived from the interpolation. The latter two are somewhat
arbitrary. Our attitude was one of being conservative enough to guarantee that the uncertainty in the
total flux was not underestimated due to the difficulty in establishing these two values.
The corresponding absolute errors were then combined to derive the errors in the flux which are shown
in Table~\ref{tab:res}. Combining these with the uncertainty in the angular diameter, we derived the
uncertainty in the individual values of the effective temperature. As a final result we take the mean
of the three values and consider the uncertainty to be the largest of the three uncertainties. Thus,
the flux and effective temperature adopted for $\gamma$ Equ are, respectively,
($3.12 \pm 0.21$)~$\times 10^{-7}$~erg~cm$^{-2}$~s$^{-1}$ and  7364~$\pm$~235~K. If, instead, we took
for the effective temperature an uncertainty such as to enclose the three uncertainties, the result
would be $T_{\rm eff}$~=~7364~$\pm$~250~K.

\begin{figure*}[t]
\centering
 \includegraphics[width=14cm, angle=0]{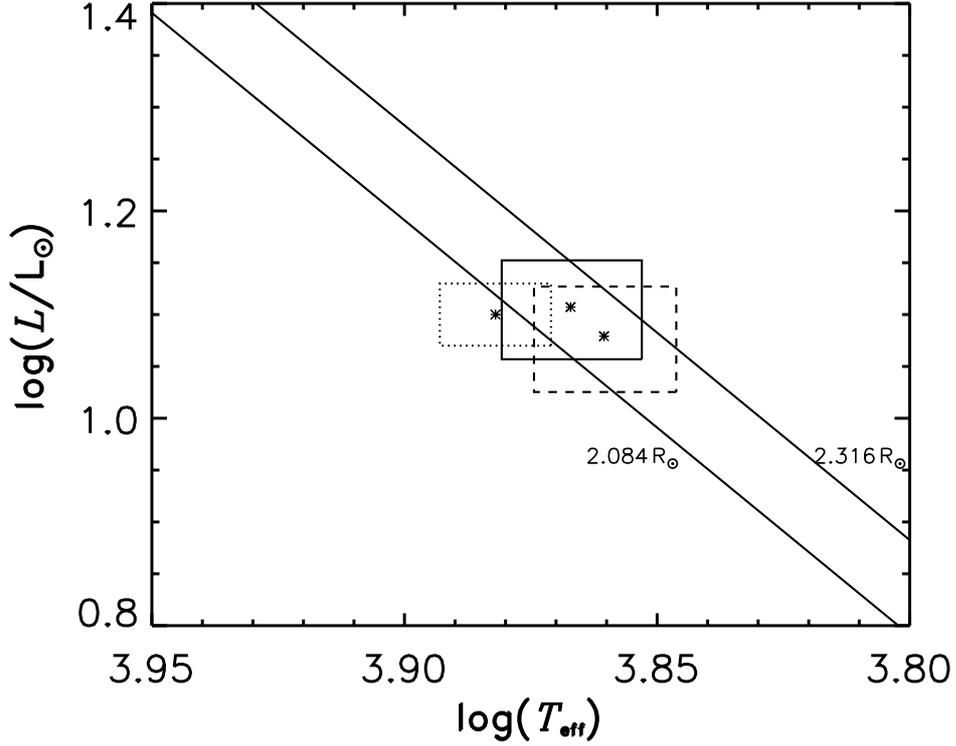}
\caption{\textbf The position of $\gamma$ Equ in the Hertzsprung-Russell diagram. The constraints on
the fundamental parameters are indicated by the 1$\sigma$-error box (log~$T_{\rm eff}$, log~($L$/L$_{\odot}$))
and the diagonal lines (radius). The box in solid lines corresponds to the results derived when ignoring the
presence of the companion star. The box in dashed lines corresponds to the results derived after subtracting
from the total bolometric flux the maximum contribution expected from the companion (see text for details).
The box in dotted lines corresponds to the fundamental parameters derived by Kochukhov \& Bagnulo (2006) and
used by Gruberbauer et al. (2008) in the asteroseismic modelling of $\gamma$ Equ.}
  \label{fig:HR}
\end{figure*}

\subsection{Contamination by the companion star}
\label{companion}

In fact, since $\gamma$ Equ is a multiple system and the distance between the primary (hereafter,
$\gamma$ Equ\,A) and the secondary (hereafter, $\gamma$ Equ\,B) is 1.25", the bolometric flux of
$\gamma$ Equ determined in Sect.~3 contains the contribution of both components. Given its magnitude,
one may anticipate that the contribution of $\gamma$ Equ\,B to the total flux will be small. Although
the data available in the literature for this component is very limited, we used them to estimate the
impact of $\gamma$ Equ\,B's contribution on our determination of the effective temperature of $\gamma$ Equ\,A.

We collected the magnitudes $m_B$~=~9.85~$\pm$~0.03 and $m_V$~=~8.69~$\pm$~0.03 of $\gamma$ Equ\,B from
Fabricius et al. (2002) and determined a value for its effective temperature using the color-$T_{\rm eff}$
calibration from \cite{ramirez05}. This was done assuming three different arbitrary values and uncertainties
for the metallicity, namely $-0.4\pm0.5$, $0\pm0.5$ and $0.4\pm0.5$ dex. The values found for the effective
temperature were $T_{\rm eff}$~=~4570, 4686 and 4833 K, respectively, with an uncertainty of $\pm 40$K
(\cite{ramirez05}). The metallicity, the effective temperature, and the absolute V-band magnitude were used to estimate
log~$g$, using theoretical isochrones from Girardi et al. (2000)\footnote{http://stev.oapd.inaf.it/cgi-bin/param}.
For the three values of metallicities and $T_{\rm eff}$ mentioned above, we found log~$g$~=~4.58, 4.53, and 4.51,
respectively. With these parameters we computed three Kurucz models and calibrated each of them in three different
ways: (i) using the H$_P=9.054\pm 0.127$ magnitude (\cite{perryman97}), (ii) using the $m_B$ magnitude, and
(iii) using the $m_V$ magnitude. To convert from Hipparcos/Tycho magnitudes into fluxes we used the zero points
from Bessel \& Castelli (private communication). The maximum flux found for $\gamma$ Equ\,B through the procedure
described above was 0.19$\times10^{-7}\, \rm {erg\, cm^{-2}\,s^{-1}}$, which corresponds to 6\% of the total flux.
This implies that the effective temperature of  $\gamma$ Equ\,A determined in the previous section may be in excess
by up to 111 K due to the contamination introduced by this companion star.

\section{Discussion}

\subsection{Position in the HR-diagram}

We derive the radius of $\gamma$ Equ thanks to the formula:
\begin{equation}
    \theta_{LD} = 9.305 * R/d,
\end{equation}
where $\theta_{LD}$ stands for the limb-darkened angular diameter (in mas),
$R$ for the stellar radius (in solar radius, ${\rm R_{\odot}}$), and $d$ for the
distance (in parsec).
We obtain $R$~=~2.20~$\pm$~0.12~${\rm R_{\odot}}$.

We use the bolometric flux $f_{bol}$ and the parallax $\pi_P$ to determine the $\gamma$ Equ's luminosity
from the relation:
\begin{equation}\label{eq3}
L = 4 \pi f_{bol} \frac{C^{2}}{{\pi_{P}}^{2}},
\end{equation}
where $C$ stands for the conversion factor from parsecs to meters.
We obtain $L/{\rm L_\odot}$~=~12.8~$\pm$~1.4 and can set $\gamma$ Equ in the HR diagram (Fig.~\ref{fig:HR}).

Recently, seismic data of $\gamma$~Equ obtained with the Canadian-led satellite MOST have been
modeled by Gruberbauer et al. (2008) based on the fundamental parameters coming from \cite{Koch06}
and using a grid of pulsation models including the effect of the magnetic field. A comparison of the
HR diagram error-box considered by the authors (see dotted-line box in Figure~\ref{fig:HR}) and
our uncertainty regions shows that the regions are considerably different. In fact, even if we do
not account for the contribution of the companion, we obtain a lower effective temperature with
log~$T_{\rm eff}$ = 3.867 $\pm$ 0.014 to be compared to log~$T_{\rm eff}$ = 3.882 $\pm$ 0.011
from Gruberbauer et al. (2008). This discrepancy between the uncertainty regions increases if the
companion contribution is taken into account. In that case, the overlap between the two regions is
very small.

As regards to luminosity, our calculation shows that for $\gamma$~Equ (as well as for $\alpha$~Cir)
the contributions of the uncertainties in the bolometric flux and parallax to the uncertainty in $L/{\rm L_\odot}$
are comparable. This is quite different from the results obtained by Kochukhov \& Bagnulo (2006)
who found that the dominant contribution to the uncertainty in $L/{\rm L_\odot}$ comes from the parallax.
The authors mentioned that the bolometric flux that was adopted in their work was that for normal stars. When
dealing with peculiar stars, like Ap stars, it may be more adequate to properly compute the bolometric flux.
However, it is precisely the difficulty in obtaining the full spectrum of the star that increases the
uncertainty in the computed bolometric flux and, hence, in the luminosity and effective temperature.
That is well illustrated by the following fact: if the somewhat arbitrary 20$\%$ uncertainties adopted in
our work for the total fluxes derived from the Kurucz model and from the interpolation, were replaced by
5$\%$ uncertainties, we would obtain formal uncertainties in $L/{\rm L_\odot}$ and $T_{\rm eff}$ comparable
and smaller, respectively, to those quoted by Kochukhov \& Bagnulo (2006).

\subsection{Bias due to stellar features}

We use the whole spectral energy density to determine the bolometric flux.
We then deduce the effective temperature from this bolometric flux and the angular diameter. The
determination of the angular diameter is based on visibility measurements that are directly linked to
the Fourier Transform of the object intensity distribution. For a single circular star, the visibility
curve as a function of spatial frequency B/$\lambda$ (where B stands for the interferometric baseline
and $\lambda$ for the operating wavelength) is related to the first Bessel function, and contains an
ever decreasing series of lobes, separated by nulls, as one observes with an increasing angular
resolution. As a rule of thumb, the first lobe of the visibility curve (see Fig.~1 for an example)
is sensitive to the size of the object only. As an example, for a star whose angular diameter equals
0.56 mas like $\gamma$ Equ, the difference in squared visibility between a uniform-disk and a limb-darkened
one is of the order of 0.5\% in the first lobe. The following lobes are sensitive to limb darkening and
atmospheric structure but consist of very low visibilities. Finally, departure from circular symmetry
(due to stellar spots from instance) requires either interferometric imaging by more than two telescopes
or measurement close to the null. As a consequence, our interferometric data collected in the first part
of the first lobe are only sensitive to the size of the target and cannot be used to study the potential
complex structure of the atmosphere.

\section{Conclusion}

Thanks to the unique capabilities of VEGA/CHARA, we present an accurate measurement of the limb-darkened
angular diameter of a target as small as 0.564 $\pm$ 0.017 mas. In combination with our estimate of the
bolometric flux based on the whole spectral energy density, we determine the effective temperature of
$\gamma$~Equ~A. Without considering the contribution of the closest companion star ($\gamma$~Equ B)
to the bolometric flux, we found an effective temperature 7364~$\pm$~235 K, which is below the effective
temperature that has been previously determined. An estimate of that contribution leads to the conclusion
that the above value may still be in excess by up to about 110 K, which increases further the discrepancy
between the literature values for the effective temperature of $\gamma$~Equ A and the value derived here.
The impact on the seismic analysis of considering the new values of the radius and effective temperature
should be considered in future modeling of this star.

More generally, this study illustrates the advantages of optical long-baseline interferometry for
providing direct and accurate angular diameter measurements and motivates observations of other
main-sequence stars to bring constraints on their evolutionary state and their internal structures.
Within this context, the operation of VEGA in the visible is very complementary to the similar
interferometric studies performed in the infrared range since it allows to study spectral types
ranging from B to late-M and thus it opens the new window of the early spectral types (\cite{vega}).

Another promising issue would be to use longer interferometric baselines to be sensitive to the stellar
spots and bring constraints on the stellar surface features.

\begin{acknowledgements}
VEGA is a collaboration between CHARA and OCA/LAOG/CRAL/LESIA that has been supported by the
French programs PNPS and ASHRA, by INSU and by the R\'egion PACA. The project has obviously
taken benefit from the strong support of the OCA and CHARA technical teams.
The CHARA Array is operated with support from the National Science
Foundation through grant AST-0908253, the W. M. Keck Foundation, the
NASA Exoplanet Science Institute, and from Georgia State University. This work was partially
supported by the projects PTDC/CTE-AST/098754/2008 and PTDC/CTE-AST/66181/2006, and the grant
SFRH / BD / 41213 / 2007 funded by FCT/MCTES, Portugal. MC is supported by a Ci{\^e}ncia 2007
contract, funded by FCT/MCTES(Portugal) and POPH/FSE (EC). This research has made use of the
SearchCal and LITPRO services of the Jean-Marie Mariotti Center, and of CDS Astronomical Databases SIMBAD and VIZIER.

\end{acknowledgements}

{}

\end{document}